\documentclass[sigconf]{acmart}
\AtBeginDocument{%
  }

\makeatletter
\@acmownedfalse
\makeatother

\acmConference[Purposeful XR '25]{Purposeful XR: Affordances, Challenges, and Speculations for an Ethical Future at CHI 2025}{April 26, 2025}{Yokohama, Japan}
\setcopyright{cc}
\setcctype{by}
\copyrightyear{2025}
\acmDOI{}
\acmISBN{}
\usepackage{xspace}
\usepackage{epigraph}

\begin{document}

\title[Change Your Perspective, Widen Your Worldview!]{Change Your Perspective, Widen Your Worldview! Societally Beneficial  Perceptual Filter Bubbles in Personalized Reality}

\author{Jannis Strecker}
\email{jannisrene.strecker@unisg.ch}
\orcid{0000-0001-7607-8064}
\affiliation{
  \institution{University of St. Gallen}
  \city{St. Gallen}
  \country{Switzerland}
}

\author{Luka Bekavac}
\email{lukajurelars.bekavac@student.unisg.ch}
\orcid{0009-0009-3598-3012}
\affiliation{%
  \institution{University of St. Gallen}
  \city{St.Gallen}
  \country{Switzerland}
}

\author{Kenan Bekta\c{s}}
\email{kenan.bektas@unisg.ch}
\orcid{0000-0003-2937-0542}
\affiliation{%
  \institution{University of St. Gallen}
  \city{St.Gallen}
  \country{Switzerland}
}

\author{Simon Mayer}
\email{simon.mayer@unisg.ch}
\orcid{0000-0001-6367-3454}
\affiliation{%
  \institution{University of St. Gallen}
  \city{St.Gallen}
  \country{Switzerland}
}

%\renewcommand{\shortauthors}{Strecker et al.}

%                       150 words
\begin{abstract}
Extended Reality (XR) technologies enable the personalized mediation of an individual's perceivable reality across modalities, thereby creating a Personalized Reality (PR). While this may lead to individually beneficial effects in the form of more efficient, more fun, and safer experiences, it may also lead to \textit{perceptual filter bubbles} since individuals are exposed predominantly or exclusively to content that is congruent with their existing beliefs and opinions. This undermining of a shared basis for interaction and discussion through constrained perceptual worldviews may impact society through increased polarization and other well-documented negative effects of filter bubbles. In this paper, we argue that this issue can be mitigated by increasing individuals' awareness of their current perspective and 
providing avenues for development, 
including through support for engineered serendipity and fostering of self-actualization that already show promise for traditional recommender systems. We discuss how these methods may be transferred to XR to yield valuable tools to give people transparency and agency over their perceptual worldviews in a responsible manner.
\end{abstract}



\maketitle

\section{Introduction}
% +++++ 4 pages max +++++++++++
% XR affordance: personalized mediation of physical reality

% societal challange: narrow, isolated view-points. perceptual filter bubbles.

% ---
% \begin{itemize}
%     \item \textbf{From submission form:}
%     \begin{itemize}
%         \item Detail your specific social challenge
%         \item Discuss how and why this challenge is currently inadequately addressed
%         \item Detail your chosen XR affordance(s) and its unique ability to address the social challenge
%         \item Analyze the technological/future advances necessary for the affordance(s) to better address the social challenge
%         \item Consider potential harms due to advancing the XR affordance(s)
%     \end{itemize}
% \end{itemize}
% ---

\epigraph{The eye forms the world / the world forms the eye.}{Marvin Hill~\cite{hill2024}}

Extended reality (XR)~\cite{rauschnabel2022} changes how we perceive and understand the world---perceptually as well as conceptually (cf.~\cite{turner2022, koltko-rivera2004}). 
XR is commonly used as an umbrella term for Virtual Reality (VR), Mixed Reality (MR)~\cite{skarbez2021}, and Augmented Reality (AR)~\cite{milgram1995}. 
%Seen broadly, XR comprises any kind of technology-mediated perception of reality across all sensory modalities (cf.\cite{skarbez2021}). 
XR experiences that mediate physical reality may augment, diminish, or substitute physical stimuli with virtual content across all sensory modalities, though vision and audio are most commonly addressed.
%from the unmodified, original reality that includes virtual and physical reality. 
Such mediation may range from small changes, such as increasing the affordance of a light switch to be ``pressable'' when a room becomes darker (e.g., as a virtual overlay), to larger changes, 
such as substituting the color scheme of everything a person perceives visually.
This puts an XR device and the entity that controls it in a powerful position, as they essentially gain the ability to, at will, ``determine how users experience the world, how they conceive of themselves, and how they regard others''~\cite[p.99]{madary2023}. 

This ability to control perception in XR contrasts with physical reality, where everyone who shares the same environment has, in principle, the possibility to perceive the same objective ground truth (given uniform sensory abilities), i.e. the unfiltered, unmediated physical environment.
Physical reality thus provides the basis for social \textit{shared worlds}~\cite{brincker2021} and \emph{intersubjectivity} (i.e. ``the common-sense, shared meanings constructed by people in their interactions with each other and used as an everyday resource to interpret the meaning of elements of social and cultural life'' ~\cite[p.1126]{seale2018}). 
However, such a communal sense-making of the world assumes that everyone involved potentially has access to the same or similar stimuli and may form congruent perspectives.
Even without any technological mediation, this assumption proves difficult, as \citeauthor{habich2019} observe that different subjective perspectives may lead to different subjective truths based on the same objective truth (e.g., the shadow of an object may be a triangle from one person's perspective and a circle from another one's)~\cite{habich2019}.
Thus, already without technological mediation, individuals have their distinct \emph{perceptual} worldview: their sensory organs allow them to perceive the environment around them from a certain perspective (think of color vision deficiencies, or discrepancies in hearing, as tangible examples). This perceptual worldview determines whether and to what extent the individual perceives stimuli, which in turn may influence its \emph{conceptual} worldview, i.e. a ``[set] of beliefs and assumptions that describe reality''~\cite[p.3]{koltko-rivera2004}. 

The dynamic mediation of a person's visually perceivable physical reality, e.g., through XR head-mounted displays (HMD), such as the Apple Vision Pro or Microsoft HoloLens 2, further complicates this assumption of a common ground. 
The philosopher David Chalmers writes: ``It’s easy to imagine that in the future, there will be multiple dominant systems of augmented reality. Instead of a single universal reality, there will be Apple Reality, Facebook Reality, and Google Reality. Each corporation will set up its own virtual worlds and augment them with its own virtual objects''~\cite[p.230]{chalmers2023}. He concludes that each of these realities will have their own objective truths that people using the other realities cannot observe. Yet, we argue that this assumption is not far-reaching enough. When considering XR systems in more detail, multiple stakeholders (e.g., device manufacturers, app developers, advertising companies) may be responsible for different parts of an XR system, such as the delivery medium and the application. The fragmentation into different objective realities might therefore be much worse, as there could exist, e.g. a ``Instagram Reality fact checked by BBC on Samsung glasses'', or even a ``TikTok Reality with ads from Google sponsored by Coca Cola on Apple glasses''. It is therefore conceivable that every person might even have multiple ``objective'' realities they could switch between.
Existing research on harmful implications of XR mentions this possible fragmentation of realities merely as a side issue, without considering it in more detail (cf.~\cite{abraham2022, eghtebas2023,rosenberg2022,turner2022}).
% Thus, the question of a common, objective ground-truth is further complicated by XR technologies.

This issue becomes even more problematic if \textit{personalization} (i.e., system-initiated interface adaption based on personal data~\cite{strecker2024a}) is added to an XR application. 
We call such personalized XR a \textit{Personalized Reality} (PR), i.e. ``a physical, virtual, or mixed reality that has been modified in response to personal user data and may be perceived by one or multiple users through any sensory modality''~\cite[p.2]{strecker2024a}.
In PR, even people who use the same application with the same device type most likely do not perceive the same virtual and physical stimuli, as each person perceives a \emph{personalized} version of reality. People thus may not be aware that their own personalized perspective is not the same or even similar to that of others. 
In consequence, their ability to interact with others and understand their environment may be diminished.
%In consequence, this may deteriorate their common interaction in and understanding of their environment.
% Thus, XR may further aggravate the fragmentation (cf.~\cite{habich2019}) of reality perception and in consequence of how people understand and interact in the world as well.

An example of where such constrained perspectives have become problematic is Web-based personalization~\cite{vesanen2007, fan2006}, such as employed by social media, music streaming, or shopping Websites.
% insert some benefits here?
Research has shown that such personalization may, e.g., lead to polarization~\cite{celis2019}, radicalization~\cite{albadi2022} and manipulation~\cite{yeung2018}.
When people use social media, for instance, the content they perceive is based on their own and others' interaction data (clicks, likes, etc.) and the platform's recommender systems.
This may situate users in \textit{filter bubbles}~\cite{pariser2011} or \textit{echo chambers}~\cite{terren2021} where the received content gradually narrows in structural, topic, and viewpoint diversity, reinforcing their existing beliefs~\cite{michiels2022}.
% The forming of filter bubbles can 
%Over time, this leads to a \emph{bubble of ideas} where users are exposed only to familiar perspectives. 
Crucially, they have little to no control over what content and ideas enter this bubble---or, possibly more importantly, what is filtered and never enters it~\cite{pariser2011}.

% connect back to PR
% motivate how we come from the issues we described to the solutions in this paper
In this paper, we argue that this is not a one-way street: the personalization aspect in PR applications itself could be exploited to show individuals that their perspective (i.e. their perceptual filter bubble) is only one possibility, that it is governed by their own choices and data, and the PR application's recommender systems. On top of such transparency, users could be empowered to actively form their perceptual filter bubble to suit their current (dynamic) goals: instead of unconsciously sliding into content filter bubbles, PR could permit users to consciously enter and leave filter bubbles while staying aware of their current information state. Conceptually, this idea is an extension of an approach we presented earlier where we focused on exchanging personalized content between people in XR to provide glimpses into each other's perceived realities~\cite{strecker2023a}. 

To this end, we discuss ways how the affordance of dynamic, personalized mediation in XR may be used to provide transparency and agency to perceptual worldviews. 
%This may help to mitigate the spreading of potential harmful implications such as polarization and radicalization to more areas of daily life.
In the remainder of the paper, we look at how the concept of \textit{filter bubbles} provides a metaphor for constrained perception in personalization systems, then we discuss approaches in recommender system research that aim at mitigating limited recommendations, and lastly, we provide examples and future avenues for how these solutions can benefit individuals and society when brought to physical reality with XR.

\section{Personalized Perspectives May Create Filter Bubbles}

More than ten years ago, Eli Pariser coined the term \emph{Filter bubbles}~\cite{pariser2011} to describe personalized digital environments (e.g., social media feeds) where algorithms restrict exposure to diverse viewpoints.
While some research has found that these may lead to polarization, and amplify systemic risks such as misinformation, manipulation, and discrimination~\cite{gordon-tapiero2022}, the term has faced criticism for lacking empirical specificity and based on inconsistent findings across studies~\cite{guess2018,terren2021}. 
% To systematically study filter bubbles, 
To enable the systematic study of filter bubbles, more recently, Michiels et al. proposed a measurable and empirically grounded definition: ``A technological filter bubble is a decrease in the diversity of a user’s recommendations over time, in any dimension of diversity, resulting from the choices made by different recommendation stakeholders''~\cite[p.275]{michiels2022}.
In addition to this technological approach, others highlight the influence of people's self-selection on filter bubbles, e.g., based on their interests or social group identity~\cite{ekstrom2022} which may lead to the formation of filter bubbles within hours of interacting with personalized recommender systems~\cite{bekavac2024a}.
People may, for instance, seek \textit{protective filter bubbles} as ``an algorithmically curated information ecosystem that shields people from threats to psychological and physical safety, including targeted threats such as hate speech, discrimination, and political persecution and generalized threats such as distressing media''~\cite[p.2]{erickson2024}. 
Thus, filter bubbles may be described as a complex interplay between technological (e.g., recommender systems), personal (e.g., one's own interactions on the platform), and social factors (e.g., friend's interactions). 
Filter bubbles are hence not problematic per se, but they do present people with a certain narrow perspective that can make it difficult to interact with others because there may not be a common ground-truth, e.g., on a particular topic.
%Filter bubbles may not necessarily be problematic per se, however, they present people with a certain, narrow perspective, that may make interactions with others difficult, as a common understanding, e.g., on a certain topic may not exist.
% Filter bubbles may arise based on the interplay of platforms' recommender systems, e.g., through exposure bias~\cite{krause2024}, and the interaction of users on that platform, e.g., through self-selection of preferred content~\cite{EKSTROM2022100226}. 

\section{Mitigating Narrow Focuses in Recommender Systems}

% In general, recommender systems build the basis for personalized systems, as they provide the algorithm that determines how the presented content for a person should be personalized (i.e. adapted based on their personal data).

In traditional recommender systems, one approach to counter the narrow focus of personalized recommendations is the inclusion of \emph{serendipity}~\cite{reviglio2023}. Serendipity, in this context, refers to the unexpected yet meaningful encounters users experience when engaging with content~\cite{smets2022}. It is a multifaceted concept composed of various elements, including unexpectedness, relevance, novelty, usefulness, and diversity, which together shape how users discover and interact with unplanned but valuable information~\cite{smets2022}. Popular examples are niche blog posts appearing in a social media feed or a unique, lesser-known product recommended on an e-commerce site. 
Such ``engineered serendipity''~\cite{Lane2021} may enrich personalized recommendations and thus mitigate filter bubbles to some extent. However, serendipitous content is, as the name implies, often random in the sense that there is little to no connection to the person's own interests or preferences. Thus, the content could be irrelevant or even harmful, e.g., when it is connected to topics the person has traumatic experiences with.

An alternative approach that presents novel and relevant content to people but in a more goal-directed, controlled way has been referred to as ``recommender systems for self-actualization'' (RSSA)~\cite{sullivan2019,knijnenburg2016}.  
\citeauthor{knijnenburg2016} define RSSA as ``personalized systems that have the explicit goal to not just present users with the best possible items, but to support users in developing, exploring, and understanding their own unique tastes and preferences''~\cite[p.1]{knijnenburg2016}.
They propose that RSSA should focus on supporting users' decision-making, exploration, and multiple tastes as opposed to \emph{replacing} their decision-making, fostering consumption, and focusing them on singular tastes.
Additionally, they suggest that RSSA could offer 'imperfect' recommendations such as ``Things we think you will hate'', ``Things we have no clue about'', or ``Things you’ll be among the first to try''\cite[p.13]{knijnenburg2016}.
Additionally, scrutable personalization systems~\cite{kay2012} are another approach that gives people more agency over their personalized content. These systems allow users to directly modify the personalization algorithm by configuring which parameters should be prioritized by the recommender system. One example of these is TikTok’s new content preference feature, where users can adjust sliders to indicate whether they want more or less of certain topics in their \emph{For You} page\footnote{\url{https://www.theverge.com/2024/8/30/24232154/tiktok-for-you-page-algorithm-content-preferencers-sliders}. Last accessed Febrauary 28, 2025}, e.g., ``creative arts,'' ``current affairs,'' or ``humor''. 
Additional approaches to enable more agency in personalization systems include multiple profiling, allowing users to create and switch between different personalized feeds; proportional opt-out, offering granular control over the balance of personalized and non-personalized recommendations; and user feedback, which has the potential to align recommendations with explicitly stated preferences. Recent regulatory advancements on online platforms, in a best-case compliance scenario, would mandate these features, ensuring that users have the right to intervene in algorithmic parameters and shape their content exposure~\cite{reviglio2024}.

We refer to a combination of these approaches to manage filter bubbles as \textit{engineered serendipity}, that is a goal-oriented recommendation of content that is outside of the current filter bubble. These goals may be defined by the affected person directly or suggested by the system based on the personal data it has.
Within the metaphor of \emph{filter bubbles}, engineered serendipity provides a method to strategically show people that they are situated in a constrained bubble, but also to manage these bubbles and ``poke holes'' in them to provide avenues for personal development.
Traditionally, personalization system rather reduce people's worldview by constraining the amount of content they perceive, engineered serendipity, in contrast, may be a method for strategically widening the scope of what people perceive. 

Few if any commercial recommender systems seem to have included effective measures to mitigate constrained perceptions. On the contrary, current engagement-based ranking approaches primarily aim to maximize attention metrics, as content creators often rely on visibility for financial survival~\cite{ovadya2023}. These incentives favor engagement-bait-sensationalized and hyperbolic content designed to provoke reactions rather than to inform. As a result, most discussions around personalized content, particularly in social media, highlight their harmful effects, including misinformation, polarization, and the reinforcement of ideological bubbles~\cite{barbera2020, terren2021, bozdag2015}.

% As XR technologies increasingly provide the means to personalize a person's full perceptual worldview, it is noteworthy or even necessary to revisit these ideas.

\section{Managing Perceptual Filter Bubbles in PR}

Currently, the impact of filter bubbles in Web-based personalization on physical reality and people's interaction in physical environments is predominantly indirect. People consume personalized content online which may influence their conceptual worldview, and thus shapes how they interact in physical reality.
PR experiences, however, may enable \textit{perceptual} filter bubbles, that directly affect how people perceive and interact in the environment they are situated in~\cite{abraham2022,strecker2023a, rosenberg2022}.
When wearing XR HMDs, people could choose to be placed in a literal perceptual bubble around them, that  e.g., selectively diminishes physical objects visually that do not align with a person's worldview ~\cite{turner2022}. 

In general, this may be beneficial, as people encounter many stimuli in physical reality, and PR applications could filter those that are most relevant to a person's current situation. 
This could enable people to gain more agency over the things they perceive, e.g., through the blocking of physical ads using XR~\cite{katins2025}, or be in a state of flow~\cite{csikszentmihalyi2014} that allows them to complete tasks more efficiently~\cite{lindlbauer2019}.

However, as we discussed, a constrained worldview may also negatively impact interactions with others because a common ground is missing. 
Existing design fictions from artists, such as ``Hyper-Reality''~\cite{matsuda2016} or ``The Lenz''~\cite{damienlutz2023}, illustrate speculatively how a PR may mediate physical reality to such an extend, that there is little connection left to the original reality and to that of others. Additionally, inspired by the episode ``Arkangel'' of the TV series Black Mirror~\footnote{\url{https://en.wikipedia.org/wiki/Arkangel_(Black_Mirror)}. Last accessed February 28, 2025.}, parents could configure the PR of their children to filter stimuli at will, suppress emotions like fear or sadness, and obscure people or objects that conflict with their own beliefs. Such a personalized, manufactured reality could severely hinder autonomy and social development, trapping children in a curated world that reinforces parental control.
Furthermore, as recommender systems are often configured to maximize monetary returns and engagement, as seen in social media~\cite{yeung2018}, it seems likely that perceptual filter bubbles in PR may be reinforced by these incentives.
% Furthermore, as monetary incentives and engagement-driven recommender algorithms dictate what information users are exposed to, as seen in social media~\cite{yeung2018}, these often reinforce filter bubbles for profits. 

Thus, we argue that any PR application should have a mechanism is place that allows to strategically manage one's perceptual filter bubble by oneself. This includes the possibility to expand the recommended content based on previously unknown perspectives, but also the option to remove perspectives one does not wish to perceive any longer. 
Additionally, the PR application could deliberately ``poke holes'' in the perceptual filter bubbles by offering ``imperfect'' recommendations, e.g., by offering a perspective contrary to the person's.
We illustrate how a PR experience making use of \textit{engineered serendipity} could look like with several examples that employ XR technology and personalization: 

\begin{itemize}
     \item \emph{News}: Websites such as Ground News\footnote{\url{https://ground.news/}. Last accessed February 23, 2025.} and AllSides\footnote{\url{https://www.allsides.com/}. Last accessed February 23, 2025.} provide news from across the political spectrum to allow people to get a balanced view on the news they consume.
    A PR could allow, e.g., right-leaning people who would like to widen their perspective towards the center or even left to make use of such an approach. When they, e.g., perceive political advertisements on a billboard, or when they are reading a (physical) newspaper or book, their PR 
    could virtually overlay information from one or more viewpoints other than their own.
    %could show virtually overlaid information from one or multiple other viewpoints that are different than their own. 
    \item \emph{Food Choices}: Our ShoppingCoach system visually diminishes food products that are not deemed healthy for a certain person based on their previous shopping history~\cite{strecker2024}. As food choice may evolve over time, an extension of our system could then gradually over the course of multiple shopping sessions shift the recommended products towards a goal the person has formulated, e.g. to buy less meat, or more regionally produced products.
   
    \item \emph{History \& Archaeology}: In a similar vein, a PR application that displays personalized XR overlays when visiting a historical or archaeological site (cf.\cite{yang2022a}), may offer to show explanations from different perspectives. Starting from the perspective the person already knows, the PR could gradually extend the presented information to connected or less connected perspectives, e.g., from underrepresented minorities, or various scholarly disciplines.
    % \item \textbf{Learning}: Language learning, gradually not replacing SOCRAR \cite{strecker2022}: gradually allow me to learn other units, plus add example of personalized language learning\cite{???}
    \item \emph{Literature}: A PR experience in a library or bookshop (cf.\cite{wei2024}) could be configured by a person to recommend them more books based on characteristics they did not consider so far.
    This could include, e.g, books from authors with origins in other cultures, or with a more (or less) diverse set of characters. Behaving `imperfectly', the PR application could suggest books the person may hate, or that no one has read so far.
    \item \emph{Navigation}: When used for personalized navigation (cf.~\cite{lee2025}), a PR application could, e.g., suggest routes that are ``off the beaten track'', routes that encourage physical exercise when a person would like to be more active, or routes that show a person a new facet of the city they live in.
    % find more examples? towards "items you might hate", "the system has no idea who likes this", "imperfect recommendations"... see Knijnenburg et al
\end{itemize}

%The examples we presented here, are not meant as an exhaustive list but rather serve as possible applications where the lens of \emph{perceptual filter bubbles} allows to highlight how these affect the interplays between perceptual and conceptual worldviews.
%
We suggest that a PR experience is in principle agnostic to beliefs and assumptions including cultural, political, and spiritual ones, and adapts to the specific person's worldview, as long as no applicable regulation is violated. However, detecting and enforcing such a violation, may prove similarly difficult as content moderation on social media~\cite{young2022}.
Furthermore, changes to one's perceptual filter bubbles may be challenging and uncomfortable. Thus it is important that any PR application is dynamically adapting and context-aware to the person's current situation which must permit self-actualization, e.g., so that people are not overwhelmed by a new perspective.
% it moderates people's perceptual worldview only in situations that allow it, e.g., so that people are not overwhelmed by a new perspective or learning objective. 
% Similarly, their current context must permit self-actualization. 
For instance, if someone is in a rush, and about to miss their train, this is likely not an appropriate time to show them some PR content to widen their worldview. Additionally, a PR application should provide different methods of how this widening of the perceptual worldview happens. This includes providing people agency and control over which topics the reality mediation may address and through which XR device and sensory modality the personalized content is delivered. 
Additionally, the system should only suggest and not force people to explore a certain new direction in a topic to not impose on their agency.
Furthermore, a PR application should be flexible enough to respect people's decision to revert back to an earlier perspective after they try another one for a certain time.

% - not paternalistic, free choice to use
% - respect if people do not want more than convenience in a given context

% \section{Future Work}

% \begin{itemize}
%     \item \textbf{V. Technological Advances and Future Research}
%     \begin{itemize}
%         \item Necessary technological advances for better addressing the social challenge
%         \item Hurdles and harms to consider
%     \end{itemize}
% \end{itemize}

\section{Conclusion and Future Work}
% \begin{itemize}
%     \item \textbf{VII. Conclusion}
%     \begin{itemize}
%         \item Recap of the social challenge and the role of XR
%         \item Summary of the unique advantages of XR affordances
%         \item Call to action for further research and implementation
%         \item Final thoughts on the potential impact of XR in the social domain
%     \end{itemize}
% \end{itemize}

In this paper, we discussed the social challenge of fragmented, constrained worldviews stemming from the use of personalization systems. We showed how XR technology and PR experiences, on one hand, may aggravate this issue by literally constraining people's perceptual worldviews, and on the other hand, we discussed approaches to mitigate these constrained perceptions, e.g., through engineered serendipity and by providing people with control and transparency options.
We discussed why it is important for individuals and society to provide these options to mitigate dystopian implications stemming from constrained perceptual worldviews.
% may therefore go either way, leading to individually and societally harmful or beneficial implications.
Thus, it is paramount for designers of PR experiences to develop these systems responsibly as we proposed before~\cite{strecker2024a} to make perceptual filter bubbles scrutable by users of PR experiences.
Towards this, PR application designers should at least consider these three questions in the design and implementation process:
(1) Is the application creating a perceptual filter bubble?
(2) Are people appropriately informed if they are potentially in a perceptual filter bubble?
(3) Do people have the means to adjust the perceptual filter bubble intuitively?

Any personalized mediation of physical reality naturally depends on technological developments, such as advances in XR device capabilities.
Currently, e.g., XR HMDs are not strong enough for daily, and especially outdoor use. Additionally, e.g., video-see through XR HMDs such as the Meta Quest Pro may lead to harmful physiological implications when worn for a longer period of time~\cite{bailenson2024}. 
Furthermore, algorithms such as object detection, and optical character recognition that are needed to detect physical content in people's environment, need to be advanced so they can run reliably on XR devices with little power consumption.
Additionally, future recommender systems need to be built so that they allow the outlined scrutiny, agency and transparency (cf.~\cite{stray2023}).

Along with the increasing proliferation of XR technologies, recommender systems and personalization, we urge researchers and practitioners to study and design PR applications in a responsible way that acknowledges the influence of the perceptual worldview on the conceptual one and vice versa. This will enable people to navigate their lives using PR experiences in an efficient, safe, and pleasurable way, while also facilitating shared worlds and common social realities.

% \begin{acks}

% \end{acks}

\bibliographystyle{ACM-Reference-Format}
\bibliography{references}


\begin{thebibliography}{51}

%%% the Web address.

\ifx \showCODEN    \undefined \def \showCODEN     #1{\unskip}     \fi
\ifx \showISBNx    \undefined \def \showISBNx     #1{\unskip}     \fi
\ifx \showISBNxiii \undefined \def \showISBNxiii  #1{\unskip}     \fi
\ifx \showISSN     \undefined \def \showISSN      #1{\unskip}     \fi
\ifx \showLCCN     \undefined \def \showLCCN      #1{\unskip}     \fi
\ifx \shownote     \undefined \def \shownote      #1{#1}          \fi
\ifx \showarticletitle \undefined \def \showarticletitle #1{#1}   \fi
\ifx \showURL      \undefined \def \showURL       {\relax}        \fi
% invisible to TeX
\providecommand\bibfield[2]{#2}
\providecommand\bibinfo[2]{#2}
\providecommand\natexlab[1]{#1}
\providecommand\showeprint[2][]{arXiv:#2}

\bibitem[Abraham et~al\mbox{.}(2022)]%
        {abraham2022}
\bibfield{author}{\bibinfo{person}{Melvin Abraham}, \bibinfo{person}{Pejman Saeghe}, \bibinfo{person}{Mark McGill}, {and} \bibinfo{person}{Mohamed Khamis}.} \bibinfo{year}{2022}\natexlab{}.
\newblock \showarticletitle{Implications of {XR} on {Privacy}, {Security} and {Behaviour}: {Insights} from {Experts}}. In \bibinfo{booktitle}{\emph{Nordic {Human}-{Computer} {Interaction} {Conference}}} \emph{(\bibinfo{series}{{NordiCHI} '22})}. \bibinfo{publisher}{Association for Computing Machinery}, \bibinfo{address}{New York, NY, USA}, \bibinfo{pages}{1--12}.
\newblock
\showISBNx{978-1-4503-9699-8}
\href{https://doi.org/10.1145/3546155.3546691}{doi:\nolinkurl{10.1145/3546155.3546691}}


\bibitem[Albadi et~al\mbox{.}(2022)]%
        {albadi2022}
\bibfield{author}{\bibinfo{person}{Nuha Albadi}, \bibinfo{person}{Maram Kurdi}, {and} \bibinfo{person}{Shivakant Mishra}.} \bibinfo{year}{2022}\natexlab{}.
\newblock \showarticletitle{Deradicalizing {YouTube}: {Characterization}, {Detection}, and {Personalization} of {Religiously} {Intolerant} {Arabic} {Videos}}.
\newblock \bibinfo{journal}{\emph{Proc. ACM Hum.-Comput. Interact.}} \bibinfo{volume}{6}, \bibinfo{number}{CSCW2} (\bibinfo{date}{Nov.} \bibinfo{year}{2022}), \bibinfo{pages}{505:1--505:25}.
\newblock
\href{https://doi.org/10.1145/3555618}{doi:\nolinkurl{10.1145/3555618}}


\bibitem[Bailenson et~al\mbox{.}(2024)]%
        {bailenson2024}
\bibfield{author}{\bibinfo{person}{Jeremy~N. Bailenson}, \bibinfo{person}{Brian Beams}, \bibinfo{person}{James Brown}, \bibinfo{person}{Cyan DeVeaux}, \bibinfo{person}{Eugy Han}, \bibinfo{person}{Anna C.~M. Queiroz}, \bibinfo{person}{Rabindra Ratan}, \bibinfo{person}{Monique Santoso}, \bibinfo{person}{Tara Srirangarajan}, \bibinfo{person}{Yujie Tao}, {and} \bibinfo{person}{Portia Wang}.} \bibinfo{year}{2024}\natexlab{}.
\newblock \showarticletitle{Seeing the world through digital prisms: {Psychological} implications of passthrough video usage in mixed reality.}
\newblock \bibinfo{journal}{\emph{Technology, Mind, and Behavior}} \bibinfo{volume}{5}, \bibinfo{number}{2} (\bibinfo{date}{June} \bibinfo{year}{2024}), \bibinfo{pages}{1--16}.
\newblock
\showISSN{2689-0208}
\href{https://doi.org/10.1037/tmb0000129}{doi:\nolinkurl{10.1037/tmb0000129}}


\bibitem[Barberá(2020)]%
        {barbera2020}
\bibfield{author}{\bibinfo{person}{Pablo Barberá}.} \bibinfo{year}{2020}\natexlab{}.
\newblock \showarticletitle{Social {Media}, {Echo} {Chambers}, and {Political} {Polarization}}.
\newblock In \bibinfo{booktitle}{\emph{Social {Media} and {Democracy}}}, \bibfield{editor}{\bibinfo{person}{Joshua~A. Tucker} {and} \bibinfo{person}{Nathaniel Persily}} (Eds.). \bibinfo{publisher}{Cambridge University Press}, \bibinfo{address}{Cambridge}, \bibinfo{pages}{34--55}.
\newblock
\showISBNx{978-1-108-83555-8}
\urldef\tempurl%
\url{https://www.cambridge.org/core/books/social-media-and-democracy/social-media-echo-chambers-and-political-polarization/333A5B4DE1B67EFF7876261118CCFE19}
\showURL{%
\tempurl}


\bibitem[Bekavac et~al\mbox{.}(2024)]%
        {bekavac2024a}
\bibfield{author}{\bibinfo{person}{Luka Bekavac}, \bibinfo{person}{Kimberly Garcia}, \bibinfo{person}{Jannis Strecker}, \bibinfo{person}{Simon Mayer}, {and} \bibinfo{person}{Aurelia Tamò-Larrieux}.} \bibinfo{year}{2024}\natexlab{}.
\newblock \showarticletitle{From {Walls} to {Windows}: {Creating} {Transparency} to {Understand} {Filter} {Bubbles} in {Social} {Media}}. In \bibinfo{booktitle}{\emph{{NORMalize} 2024: {The} {Second} {Workshop} on the {Normative} {Design} and {Evaluation} of {Recommender} {Systems}, co-located with the {ACM} {Conference} on {Recommender} {Systems} 2024 ({RecSys} 2024)}}. \bibinfo{publisher}{CEUR Workshop Proceedings}, \bibinfo{address}{Bari, Italy}, \bibinfo{pages}{12}.
\newblock
\urldef\tempurl%
\url{https://ceur-ws.org/Vol-3898/paper2.pdf}
\showURL{%
\tempurl}


\bibitem[Bozdag and van~den Hoven(2015)]%
        {bozdag2015}
\bibfield{author}{\bibinfo{person}{Engin Bozdag} {and} \bibinfo{person}{Jeroen van~den Hoven}.} \bibinfo{year}{2015}\natexlab{}.
\newblock \showarticletitle{Breaking the filter bubble: democracy and design}.
\newblock \bibinfo{journal}{\emph{Ethics and Information Technology}} \bibinfo{volume}{17}, \bibinfo{number}{4} (\bibinfo{date}{Dec.} \bibinfo{year}{2015}), \bibinfo{pages}{249--265}.
\newblock
\showISSN{1388-1957, 1572-8439}
\href{https://doi.org/10.1007/s10676-015-9380-y}{doi:\nolinkurl{10.1007/s10676-015-9380-y}}


\bibitem[Brincker(2021)]%
        {brincker2021}
\bibfield{author}{\bibinfo{person}{Maria Brincker}.} \bibinfo{year}{2021}\natexlab{}.
\newblock \showarticletitle{Disoriented and {Alone} in the “{Experience} {Machine}” – {On} {Netflix}, {Shared} {World} {Deceptions} and the {Consequences} of {Deepening} {Algorithmic} {Personalization}}.
\newblock \bibinfo{journal}{\emph{SATS}} \bibinfo{volume}{22}, \bibinfo{number}{1} (\bibinfo{date}{July} \bibinfo{year}{2021}), \bibinfo{pages}{75--96}.
\newblock
\showISSN{1869-7577}
\href{https://doi.org/10.1515/sats-2021-0005}{doi:\nolinkurl{10.1515/sats-2021-0005}}


\bibitem[Celis et~al\mbox{.}(2019)]%
        {celis2019}
\bibfield{author}{\bibinfo{person}{L.~Elisa Celis}, \bibinfo{person}{Sayash Kapoor}, \bibinfo{person}{Farnood Salehi}, {and} \bibinfo{person}{Nisheeth Vishnoi}.} \bibinfo{year}{2019}\natexlab{}.
\newblock \showarticletitle{Controlling {Polarization} in {Personalization}: {An} {Algorithmic} {Framework}}. In \bibinfo{booktitle}{\emph{Proceedings of the {Conference} on {Fairness}, {Accountability}, and {Transparency}}}. \bibinfo{publisher}{ACM}, \bibinfo{address}{Atlanta, GA, USA}, \bibinfo{pages}{160--169}.
\newblock
\showISBNx{978-1-4503-6125-5}
\href{https://doi.org/10.1145/3287560.3287601}{doi:\nolinkurl{10.1145/3287560.3287601}}


\bibitem[Chalmers(2023)]%
        {chalmers2023}
\bibfield{author}{\bibinfo{person}{David~J. Chalmers}.} \bibinfo{year}{2023}\natexlab{}.
\newblock \bibinfo{booktitle}{\emph{Reality+: {Virtual} {Worlds} and the {Problems} of {Philosophy}}}.
\newblock \bibinfo{publisher}{Penguin Books Ltd}, \bibinfo{address}{S.l.}
\newblock
\showISBNx{978-0-14-198678-4}


\bibitem[Csikszentmihalyi(2014)]%
        {csikszentmihalyi2014}
\bibfield{author}{\bibinfo{person}{Mihaly Csikszentmihalyi}.} \bibinfo{year}{2014}\natexlab{}.
\newblock \showarticletitle{Toward a psychology of optimal experience}.
\newblock In \bibinfo{booktitle}{\emph{Flow and the foundations of positive psychology}}. \bibinfo{publisher}{Springer}, \bibinfo{pages}{209--226}.
\newblock


\bibitem[Eghtebas et~al\mbox{.}(2023)]%
        {eghtebas2023}
\bibfield{author}{\bibinfo{person}{Chloe Eghtebas}, \bibinfo{person}{Gudrun Klinker}, \bibinfo{person}{Susanne Boll}, {and} \bibinfo{person}{Marion Koelle}.} \bibinfo{year}{2023}\natexlab{}.
\newblock \showarticletitle{Co-{Speculating} on {Dark} {Scenarios} and {Unintended} {Consequences} of a {Ubiquitous}(ly) {Augmented} {Reality}}. In \bibinfo{booktitle}{\emph{Proceedings of the 2023 {ACM} {Designing} {Interactive} {Systems} {Conference}}}. \bibinfo{publisher}{ACM}, \bibinfo{address}{Pittsburgh PA USA}, \bibinfo{pages}{2392--2407}.
\newblock
\showISBNx{978-1-4503-9893-0}
\href{https://doi.org/10.1145/3563657.3596073}{doi:\nolinkurl{10.1145/3563657.3596073}}


\bibitem[Ekström et~al\mbox{.}(2022)]%
        {ekstrom2022}
\bibfield{author}{\bibinfo{person}{Axel~G. Ekström}, \bibinfo{person}{Diederick~C. Niehorster}, {and} \bibinfo{person}{Erik~J. Olsson}.} \bibinfo{year}{2022}\natexlab{}.
\newblock \showarticletitle{Self-imposed filter bubbles: {Selective} attention and exposure in online search}.
\newblock \bibinfo{journal}{\emph{Computers in Human Behavior Reports}}  \bibinfo{volume}{7} (\bibinfo{date}{Aug.} \bibinfo{year}{2022}), \bibinfo{pages}{100226}.
\newblock
\showISSN{24519588}
\href{https://doi.org/10.1016/j.chbr.2022.100226}{doi:\nolinkurl{10.1016/j.chbr.2022.100226}}


\bibitem[Erickson(2024)]%
        {erickson2024}
\bibfield{author}{\bibinfo{person}{Jacob Erickson}.} \bibinfo{year}{2024}\natexlab{}.
\newblock \showarticletitle{Rethinking the filter bubble? {Developing} a research agenda for the protective filter bubble}.
\newblock \bibinfo{journal}{\emph{Big Data \& Society}} \bibinfo{volume}{11}, \bibinfo{number}{1} (\bibinfo{date}{March} \bibinfo{year}{2024}), \bibinfo{pages}{20539517241231276}.
\newblock
\showISSN{2053-9517}
\href{https://doi.org/10.1177/20539517241231276}{doi:\nolinkurl{10.1177/20539517241231276}}


\bibitem[Fan and Poole(2006)]%
        {fan2006}
\bibfield{author}{\bibinfo{person}{Haiyan Fan} {and} \bibinfo{person}{Marshall~Scott Poole}.} \bibinfo{year}{2006}\natexlab{}.
\newblock \showarticletitle{What {Is} {Personalization}? {Perspectives} on the {Design} and {Implementation} of {Personalization} in {Information} {Systems}}.
\newblock \bibinfo{journal}{\emph{Journal of Organizational Computing and Electronic Commerce}} \bibinfo{volume}{16}, \bibinfo{number}{3} (\bibinfo{year}{2006}), \bibinfo{pages}{179--202}.
\newblock
\showISSN{1091-9392}
\href{https://doi.org/10.1207/s15327744joce1603&4_2}{doi:\nolinkurl{10.1207/s15327744joce1603&4_2}}


\bibitem[Gordon-Tapiero et~al\mbox{.}(2022)]%
        {gordon-tapiero2022}
\bibfield{author}{\bibinfo{person}{Ayelet Gordon-Tapiero}, \bibinfo{person}{Alexandra Wood}, {and} \bibinfo{person}{Katrina Ligett}.} \bibinfo{year}{2022}\natexlab{}.
\newblock \showarticletitle{The {Case} for {Establishing} a {Collective} {Perspective} to {Address} the {Harms} of {Platform} {Personalization}}. In \bibinfo{booktitle}{\emph{Proceedings of the 2022 {Symposium} on {Computer} {Science} and {Law}}}. \bibinfo{publisher}{ACM}, \bibinfo{address}{Washington DC USA}, \bibinfo{pages}{119--130}.
\newblock
\showISBNx{978-1-4503-9234-1}
\href{https://doi.org/10.1145/3511265.3550450}{doi:\nolinkurl{10.1145/3511265.3550450}}


\bibitem[Guess et~al\mbox{.}(2018)]%
        {guess2018}
\bibfield{author}{\bibinfo{person}{Andrew Guess}, \bibinfo{person}{Brendan Nyhan}, \bibinfo{person}{Benjamin Lyons}, {and} \bibinfo{person}{Jason Reifler}.} \bibinfo{year}{2018}\natexlab{}.
\newblock \bibinfo{booktitle}{\emph{Why selective exposure to like-minded political news is less prevalent than you think}}.
\newblock \bibinfo{type}{{T}echnical {R}eport}. \bibinfo{institution}{Knight Foundation}. \bibinfo{pages}{26} pages.
\newblock


\bibitem[Habich and Nowotny(2019)]%
        {habich2019}
\bibfield{author}{\bibinfo{person}{Jörg Habich} {and} \bibinfo{person}{Verena Nowotny}.} \bibinfo{year}{2019}\natexlab{}.
\newblock \bibinfo{booktitle}{\emph{Fragmented {Realities} – {Searching} for a {Common} {Understanding} of {Truth}}}.
\newblock \bibinfo{type}{{T}echnical {R}eport}. \bibinfo{institution}{Bertelsmann Stiftung}, \bibinfo{address}{Salzburg}. \bibinfo{pages}{17} pages.
\newblock


\bibitem[Hill(2024)]%
        {hill2024}
\bibfield{author}{\bibinfo{person}{Martin Hill}.} \bibinfo{year}{2024}\natexlab{}.
\newblock \bibinfo{title}{"{Eye} {Forms} the {World}, the {World} {Forms} the {Eye}, {The}"}.
\newblock
\urldef\tempurl%
\url{http://www.marvinhill.com/product/1018}
\showURL{%
\tempurl}


\bibitem[Katins et~al\mbox{.}(2025)]%
        {katins2025}
\bibfield{author}{\bibinfo{person}{Christopher Katins}, \bibinfo{person}{Jannis Strecker}, \bibinfo{person}{Jan Hinrichs}, \bibinfo{person}{Pascal Knierim}, \bibinfo{person}{Bastian Pfleging}, {and} \bibinfo{person}{Thomas Kosch}.} \bibinfo{year}{2025}\natexlab{}.
\newblock \showarticletitle{Ad-{Blocked} {Reality}: {Evaluating} {User} {Perceptions} of {Content} {Blocking} {Concepts} {Using} {Extended} {Reality}}. In \bibinfo{booktitle}{\emph{{CHI} {Conference} on {Human} {Factors} in {Computing} {Systems} ({CHI} ’25)}}. \bibinfo{publisher}{ACM}, \bibinfo{address}{New York, NY, USA}, \bibinfo{pages}{18}.
\newblock
\showISBNx{979-8-4007-1394-1}
\href{https://doi.org/10.1145/3706598.3713230}{doi:\nolinkurl{10.1145/3706598.3713230}}


\bibitem[Kay and Kummerfeld(2012)]%
        {kay2012}
\bibfield{author}{\bibinfo{person}{Judy Kay} {and} \bibinfo{person}{Bob Kummerfeld}.} \bibinfo{year}{2012}\natexlab{}.
\newblock \showarticletitle{Creating personalized systems that people can scrutinize and control: {Drivers}, principles and experience}.
\newblock \bibinfo{journal}{\emph{ACM Transactions on Interactive Intelligent Systems}} \bibinfo{volume}{2}, \bibinfo{number}{4} (\bibinfo{date}{Dec.} \bibinfo{year}{2012}), \bibinfo{pages}{1--42}.
\newblock
\showISSN{2160-6455, 2160-6463}
\href{https://doi.org/10.1145/2395123.2395129}{doi:\nolinkurl{10.1145/2395123.2395129}}


\bibitem[Knijnenburg et~al\mbox{.}(2016)]%
        {knijnenburg2016}
\bibfield{author}{\bibinfo{person}{Bart~P. Knijnenburg}, \bibinfo{person}{Saadhika Sivakumar}, {and} \bibinfo{person}{Daricia Wilkinson}.} \bibinfo{year}{2016}\natexlab{}.
\newblock \showarticletitle{Recommender {Systems} for {Self}-{Actualization}}. In \bibinfo{booktitle}{\emph{Proceedings of the 10th {ACM} {Conference} on {Recommender} {Systems}}} \emph{(\bibinfo{series}{{RecSys} '16})}. \bibinfo{publisher}{Association for Computing Machinery}, \bibinfo{address}{New York, NY, USA}, \bibinfo{pages}{11--14}.
\newblock
\showISBNx{978-1-4503-4035-9}
\href{https://doi.org/10.1145/2959100.2959189}{doi:\nolinkurl{10.1145/2959100.2959189}}


\bibitem[Koltko-Rivera(2004)]%
        {koltko-rivera2004}
\bibfield{author}{\bibinfo{person}{Mark~E. Koltko-Rivera}.} \bibinfo{year}{2004}\natexlab{}.
\newblock \showarticletitle{The {Psychology} of {Worldviews}}.
\newblock \bibinfo{journal}{\emph{Review of General Psychology}} \bibinfo{volume}{8}, \bibinfo{number}{1} (\bibinfo{date}{March} \bibinfo{year}{2004}), \bibinfo{pages}{3--58}.
\newblock
\showISSN{1089-2680}
\href{https://doi.org/10.1037/1089-2680.8.1.3}{doi:\nolinkurl{10.1037/1089-2680.8.1.3}}
\newblock
\shownote{Publisher: SAGE Publications Inc}.


\bibitem[Lane et~al\mbox{.}(2021)]%
        {Lane2021}
\bibfield{author}{\bibinfo{person}{Jacqueline~N. Lane}, \bibinfo{person}{Ina Ganguli}, \bibinfo{person}{Patrick Gaule}, \bibinfo{person}{Eva Guinan}, {and} \bibinfo{person}{Karim~R. Lakhani}.} \bibinfo{year}{2021}\natexlab{}.
\newblock \showarticletitle{Engineering serendipity: {When} does knowledge sharing lead to knowledge production?}
\newblock \bibinfo{journal}{\emph{Strategic Management Journal}} \bibinfo{volume}{42}, \bibinfo{number}{6} (\bibinfo{year}{2021}), \bibinfo{pages}{1215--1244}.
\newblock
\showISSN{1097-0266}
\href{https://doi.org/10.1002/smj.3256}{doi:\nolinkurl{10.1002/smj.3256}}


\bibitem[Lee and Stuerzlinger(2025)]%
        {lee2025}
\bibfield{author}{\bibinfo{person}{Jong-in Lee} {and} \bibinfo{person}{Wolfgang Stuerzlinger}.} \bibinfo{year}{2025}\natexlab{}.
\newblock \showarticletitle{Towards {Personalized} {Navigation} in {XR}: {Design} {Recommendations} to {Accommodate} {Individual} {Differences}}. In \bibinfo{booktitle}{\emph{3rd {Workshop} on {Locomotion} and {Wayfinding} in {XR} at {IEEE} {VR} 2025}}. \bibinfo{address}{Saint Malo, France}, \bibinfo{pages}{5}.
\newblock


\bibitem[Lindlbauer et~al\mbox{.}(2019)]%
        {lindlbauer2019}
\bibfield{author}{\bibinfo{person}{David Lindlbauer}, \bibinfo{person}{Anna~Maria Feit}, {and} \bibinfo{person}{Otmar Hilliges}.} \bibinfo{year}{2019}\natexlab{}.
\newblock \showarticletitle{Context-{Aware} {Online} {Adaptation} of {Mixed} {Reality} {Interfaces}}. In \bibinfo{booktitle}{\emph{Proceedings of the 32nd {Annual} {ACM} {Symposium} on {User} {Interface} {Software} and {Technology}}}. \bibinfo{publisher}{ACM}, \bibinfo{address}{New York, NY, USA}, \bibinfo{pages}{147--160}.
\newblock
\showISBNx{978-1-4503-6816-2}
\href{https://doi.org/10.1145/3332165.3347945}{doi:\nolinkurl{10.1145/3332165.3347945}}


\bibitem[Lutz(2023)]%
        {damienlutz2023}
\bibfield{author}{\bibinfo{person}{Damien Lutz}.} \bibinfo{year}{2023}\natexlab{}.
\newblock \bibinfo{title}{The {Maya} {Lenz} - {A} speculative design concept for future {AR}}.
\newblock
\urldef\tempurl%
\url{https://www.damienlutz.com.au/thelenz/}
\showURL{%
\tempurl}
\newblock
\shownote{Retrieved June 08, 2023}.


\bibitem[Madary(2023)]%
        {madary2023}
\bibfield{author}{\bibinfo{person}{Michael Madary}.} \bibinfo{year}{2023}\natexlab{}.
\newblock \showarticletitle{Mediated {Reality}}.
\newblock In \bibinfo{booktitle}{\emph{Exploring {Extended} {Realities}}}. \bibinfo{publisher}{Routledge}, \bibinfo{pages}{19}.
\newblock
\showISBNx{978-1-003-35949-4}


\bibitem[Matsuda(2016)]%
        {matsuda2016}
\bibfield{author}{\bibinfo{person}{Keiichi Matsuda}.} \bibinfo{year}{2016}\natexlab{}.
\newblock \bibinfo{title}{Hyper-{Reality}}.
\newblock
\urldef\tempurl%
\url{http://hyper-reality.co/}
\showURL{%
\tempurl}
\newblock
\shownote{Retrieved November 21, 2023}.


\bibitem[Michiels et~al\mbox{.}(2022)]%
        {michiels2022}
\bibfield{author}{\bibinfo{person}{Lien Michiels}, \bibinfo{person}{Jens Leysen}, \bibinfo{person}{Annelien Smets}, {and} \bibinfo{person}{Bart Goethals}.} \bibinfo{year}{2022}\natexlab{}.
\newblock \showarticletitle{What {Are} {Filter} {Bubbles} {Really}? {A} {Review} of the {Conceptual} and {Empirical} {Work}}. In \bibinfo{booktitle}{\emph{Adjunct {Proceedings} of the 30th {ACM} {Conference} on {User} {Modeling}, {Adaptation} and {Personalization}}} \emph{(\bibinfo{series}{{UMAP} '22 {Adjunct}})}. \bibinfo{publisher}{Association for Computing Machinery}, \bibinfo{address}{New York, NY, USA}, \bibinfo{pages}{274--279}.
\newblock
\showISBNx{978-1-4503-9232-7}
\href{https://doi.org/10.1145/3511047.3538028}{doi:\nolinkurl{10.1145/3511047.3538028}}


\bibitem[Milgram et~al\mbox{.}(1995)]%
        {milgram1995}
\bibfield{author}{\bibinfo{person}{Paul Milgram}, \bibinfo{person}{Haruo Takemura}, \bibinfo{person}{Akira Utsumi}, {and} \bibinfo{person}{Fumio Kishino}.} \bibinfo{year}{1995}\natexlab{}.
\newblock \showarticletitle{Augmented reality: a class of displays on the reality-virtuality continuum}. In \bibinfo{booktitle}{\emph{Proc. {SPIE} 2351}}. \bibinfo{publisher}{SPIE}, \bibinfo{pages}{282--292}.
\newblock
\href{https://doi.org/10.1117/12.197321}{doi:\nolinkurl{10.1117/12.197321}}


\bibitem[Ovadya and Thorburn(2023)]%
        {ovadya2023}
\bibfield{author}{\bibinfo{person}{Aviv Ovadya} {and} \bibinfo{person}{Luke Thorburn}.} \bibinfo{year}{2023}\natexlab{}.
\newblock \bibinfo{title}{Bridging {Systems}: {Open} {Problems} for {Countering} {Destructive} {Divisiveness} across {Ranking}, {Recommenders}, and {Governance}}.
\newblock
\urldef\tempurl%
\url{http://arxiv.org/abs/2301.09976}
\showURL{%
\tempurl}
\newblock
\shownote{arXiv:2301.09976 [cs]}.


\bibitem[Pariser(2011)]%
        {pariser2011}
\bibfield{author}{\bibinfo{person}{Eli Pariser}.} \bibinfo{year}{2011}\natexlab{}.
\newblock \bibinfo{booktitle}{\emph{The {Filter} {Bubble}: {How} the {New} {Personalized} {Web} {Is} {Changing} {What} {We} {Read} and {How} {We} {Think}}}.
\newblock \bibinfo{publisher}{Penguin Books}, \bibinfo{address}{New York, NY, USA}.
\newblock
\showISBNx{978-1-101-51512-9}


\bibitem[Rauschnabel et~al\mbox{.}(2022)]%
        {rauschnabel2022}
\bibfield{author}{\bibinfo{person}{Philipp~A. Rauschnabel}, \bibinfo{person}{Reto Felix}, \bibinfo{person}{Chris Hinsch}, \bibinfo{person}{Hamza Shahab}, {and} \bibinfo{person}{Florian Alt}.} \bibinfo{year}{2022}\natexlab{}.
\newblock \showarticletitle{What is {XR}? {Towards} a {Framework} for {Augmented} and {Virtual} {Reality}}.
\newblock \bibinfo{journal}{\emph{Computers in Human Behavior}}  \bibinfo{volume}{133} (\bibinfo{date}{Aug.} \bibinfo{year}{2022}), \bibinfo{pages}{107289}.
\newblock
\showISSN{0747-5632}
\href{https://doi.org/10.1016/j.chb.2022.107289}{doi:\nolinkurl{10.1016/j.chb.2022.107289}}


\bibitem[Reviglio(2023)]%
        {reviglio2023}
\bibfield{author}{\bibinfo{person}{Urbano Reviglio}.} \bibinfo{year}{2023}\natexlab{}.
\newblock \showarticletitle{Serendipity as a {Design} {Principle} of {Personalization} {Systems}—{Theoretical} {Distinctions}}.
\newblock In \bibinfo{booktitle}{\emph{Serendipity {Science}}}, \bibfield{editor}{\bibinfo{person}{Samantha Copeland}, \bibinfo{person}{Wendy Ross}, {and} \bibinfo{person}{Martin Sand}} (Eds.). \bibinfo{publisher}{Springer International Publishing}, \bibinfo{address}{Cham}, \bibinfo{pages}{145--165}.
\newblock
\showISBNx{978-3-031-33528-0 978-3-031-33529-7}
\href{https://doi.org/10.1007/978-3-031-33529-7_8}{doi:\nolinkurl{10.1007/978-3-031-33529-7_8}}


\bibitem[Reviglio and Fabbri(2024)]%
        {reviglio2024}
\bibfield{author}{\bibinfo{person}{Urbano Reviglio} {and} \bibinfo{person}{Matteo Fabbri}.} \bibinfo{year}{2024}\natexlab{}.
\newblock \showarticletitle{Navigating the {Digital} {Services} {Act}: {Scenarios} of transparency and user control in {VLOPs}' recommender systems}. In \bibinfo{booktitle}{\emph{{NORMalize} 2024: {The} {Second} {Workshop} on the {Normative} {Design} and {Evaluation} of {Recommender} {Systems}, co-located with the {ACM} {Conference} on {Recommender} {Systems} 2024 ({RecSys} 2024)}}. \bibinfo{address}{Bari Italy}, \bibinfo{pages}{9}.
\newblock
\href{https://doi.org/10.2139/ssrn.5040307}{doi:\nolinkurl{10.2139/ssrn.5040307}}


\bibitem[Rosenberg(2022)]%
        {rosenberg2022}
\bibfield{author}{\bibinfo{person}{Louis~B. Rosenberg}.} \bibinfo{year}{2022}\natexlab{}.
\newblock \showarticletitle{Augmented {Reality}: {Reflections} at {Thirty} {Years}}. In \bibinfo{booktitle}{\emph{Proceedings of the {Future} {Technologies} {Conference} ({FTC}) 2021, {Volume} 1}} \emph{(\bibinfo{series}{Lecture {Notes} in {Networks} and {Systems}})}, \bibfield{editor}{\bibinfo{person}{Kohei Arai}} (Ed.). \bibinfo{publisher}{Springer International Publishing}, \bibinfo{address}{Cham}, \bibinfo{pages}{1--11}.
\newblock
\showISBNx{978-3-030-89906-6}
\href{https://doi.org/10.1007/978-3-030-89906-6_1}{doi:\nolinkurl{10.1007/978-3-030-89906-6_1}}


\bibitem[Seale(2018)]%
        {seale2018}
\bibfield{editor}{\bibinfo{person}{Clive Seale}} (Ed.). \bibinfo{year}{2018}\natexlab{}.
\newblock \bibinfo{booktitle}{\emph{Researching society and culture} (\bibinfo{edition}{4th edition} ed.)}.
\newblock \bibinfo{publisher}{Sage}, \bibinfo{address}{Los Angeles; London}.
\newblock
\showISBNx{978-1-4739-4716-0 978-1-4739-4715-3}


\bibitem[Skarbez et~al\mbox{.}(2021)]%
        {skarbez2021}
\bibfield{author}{\bibinfo{person}{Richard Skarbez}, \bibinfo{person}{Missie Smith}, {and} \bibinfo{person}{Mary~C. Whitton}.} \bibinfo{year}{2021}\natexlab{}.
\newblock \showarticletitle{Revisiting {Milgram} and {Kishino}'s {Reality}-{Virtuality} {Continuum}}.
\newblock \bibinfo{journal}{\emph{Frontiers in Virtual Reality}}  \bibinfo{volume}{2} (\bibinfo{date}{March} \bibinfo{year}{2021}), \bibinfo{pages}{647997}.
\newblock
\showISSN{2673-4192}
\href{https://doi.org/10.3389/frvir.2021.647997}{doi:\nolinkurl{10.3389/frvir.2021.647997}}


\bibitem[Smets et~al\mbox{.}(2022)]%
        {smets2022}
\bibfield{author}{\bibinfo{person}{Annelien Smets}, \bibinfo{person}{Lien Michiels}, \bibinfo{person}{Toine Bogers}, {and} \bibinfo{person}{Lennart Björneborn}.} \bibinfo{year}{2022}\natexlab{}.
\newblock \showarticletitle{Serendipity in {Recommender} {Systems} {Beyond} the {Algorithm}: {A} {Feature} {Repository} and {Experimental} {Design}}. In \bibinfo{booktitle}{\emph{Proceedings of the 9th {Joint} {Workshop} on {Interfaces} and {Human} {Decision} {Making} for {Recommender} {Systems}}}, Vol.~\bibinfo{volume}{Vol-3222}. \bibinfo{publisher}{CEUR Workshop Proceedings}, \bibinfo{pages}{21}.
\newblock


\bibitem[Stray et~al\mbox{.}(2023)]%
        {stray2023}
\bibfield{author}{\bibinfo{person}{Jonathan Stray}, \bibinfo{person}{Alon Halevy}, \bibinfo{person}{Parisa Assar}, \bibinfo{person}{Dylan Hadfield-Menell}, \bibinfo{person}{Craig Boutilier}, \bibinfo{person}{Amar Ashar}, \bibinfo{person}{Chloe Bakalar}, \bibinfo{person}{Lex Beattie}, \bibinfo{person}{Michael Ekstrand}, \bibinfo{person}{Claire Leibowicz}, \bibinfo{person}{Connie Moon~Sehat}, \bibinfo{person}{Sara Johansen}, \bibinfo{person}{Lianne Kerlin}, \bibinfo{person}{David Vickrey}, \bibinfo{person}{Spandana Singh}, \bibinfo{person}{Sanne Vrijenhoek}, \bibinfo{person}{Amy Zhang}, \bibinfo{person}{Mckane Andrus}, \bibinfo{person}{Natali Helberger}, \bibinfo{person}{Polina Proutskova}, \bibinfo{person}{Tanushree Mitra}, {and} \bibinfo{person}{Nina Vasan}.} \bibinfo{year}{2023}\natexlab{}.
\newblock \showarticletitle{Building {Human} {Values} into {Recommender} {Systems}: {An} {Interdisciplinary} {Synthesis}}.
\newblock \bibinfo{journal}{\emph{ACM Transactions on Recommender Systems}}  \bibinfo{volume}{2} (\bibinfo{date}{Nov.} \bibinfo{year}{2023}), \bibinfo{pages}{62}.
\newblock
\href{https://doi.org/10.1145/3632297}{doi:\nolinkurl{10.1145/3632297}}


\bibitem[Strecker et~al\mbox{.}(2024a)]%
        {strecker2024a}
\bibfield{author}{\bibinfo{person}{Jannis Strecker}, \bibinfo{person}{Simon Mayer}, {and} \bibinfo{person}{Kenan Bektas}.} \bibinfo{year}{2024}\natexlab{a}.
\newblock \showarticletitle{Personalized {Reality}: {Challenges} of {Responsible} {Ubiquitous} {Personalization}}. In \bibinfo{booktitle}{\emph{Mensch und {Computer} 2024 - {Workshopband}}}. \bibinfo{publisher}{Gesellschaft für Informatik e.V.}, \bibinfo{address}{Karlsruhe, Germany}, \bibinfo{pages}{5}.
\newblock
\href{https://doi.org/10.18420/muc2024-mci-ws11-200}{doi:\nolinkurl{10.18420/muc2024-mci-ws11-200}}


\bibitem[Strecker et~al\mbox{.}(2023)]%
        {strecker2023a}
\bibfield{author}{\bibinfo{person}{Jannis Strecker}, \bibinfo{person}{Simon Mayer}, {and} \bibinfo{person}{Kenan Bektaş}.} \bibinfo{year}{2023}\natexlab{}.
\newblock \showarticletitle{Sharing {Personalized} {Mixed} {Reality} {Experiences}}. In \bibinfo{booktitle}{\emph{Mensch und {Computer} 2023 – {Workshopband}}}. \bibinfo{publisher}{Gesellschaft für Informatik e.V.}, \bibinfo{address}{Rapperswil, SG}, \bibinfo{pages}{3}.
\newblock
\href{https://doi.org/10.18420/muc2023-mci-ws12-263}{doi:\nolinkurl{10.18420/muc2023-mci-ws12-263}}


\bibitem[Strecker et~al\mbox{.}(2024b)]%
        {strecker2024}
\bibfield{author}{\bibinfo{person}{Jannis Strecker}, \bibinfo{person}{Jing Wu}, \bibinfo{person}{Kenan Bektaş}, \bibinfo{person}{Conrad Vaslin}, {and} \bibinfo{person}{Simon Mayer}.} \bibinfo{year}{2024}\natexlab{b}.
\newblock \showarticletitle{{ShoppingCoach}: {Using} {Diminished} {Reality} to {Prevent} {Unhealthy} {Food} {Choices} in an {Offline} {Supermarket} {Scenario}}. In \bibinfo{booktitle}{\emph{Extended {Abstracts} of the 2024 {CHI} {Conference} on {Human} {Factors} in {Computing} {Systems}}} \emph{(\bibinfo{series}{{CHI} {EA} '24})}. \bibinfo{publisher}{Association for Computing Machinery}, \bibinfo{address}{New York, NY, USA}, \bibinfo{pages}{1--8}.
\newblock
\showISBNx{979-8-4007-0331-7}
\href{https://doi.org/10.1145/3613905.3650795}{doi:\nolinkurl{10.1145/3613905.3650795}}


\bibitem[Sullivan et~al\mbox{.}(2019)]%
        {sullivan2019}
\bibfield{author}{\bibinfo{person}{Emily Sullivan}, \bibinfo{person}{Dimitrios Bountouridis}, \bibinfo{person}{Jaron Harambam}, \bibinfo{person}{Shabnam Najafian}, \bibinfo{person}{Felicia Loecherbach}, \bibinfo{person}{Mykola Makhortykh}, \bibinfo{person}{Domokos Kelen}, \bibinfo{person}{Daricia Wilkinson}, \bibinfo{person}{David Graus}, {and} \bibinfo{person}{Nava Tintarev}.} \bibinfo{year}{2019}\natexlab{}.
\newblock \showarticletitle{Reading {News} with a {Purpose}: {Explaining} {User} {Profiles} for {Self}-{Actualization}}. In \bibinfo{booktitle}{\emph{Adjunct {Publication} of the 27th {Conference} on {User} {Modeling}, {Adaptation} and {Personalization}}}. \bibinfo{publisher}{ACM}, \bibinfo{address}{Larnaca Cyprus}, \bibinfo{pages}{241--245}.
\newblock
\showISBNx{978-1-4503-6711-0}
\href{https://doi.org/10.1145/3314183.3323456}{doi:\nolinkurl{10.1145/3314183.3323456}}


\bibitem[Terren and Borge-Bravo(2021)]%
        {terren2021}
\bibfield{author}{\bibinfo{person}{Ludovic Terren} {and} \bibinfo{person}{Rosa Borge-Bravo}.} \bibinfo{year}{2021}\natexlab{}.
\newblock \showarticletitle{Echo {Chambers} on {Social} {Media}: {A} {Systematic} {Review} of the {Literature}}.
\newblock \bibinfo{journal}{\emph{Review of Communication Research}}  \bibinfo{volume}{9} (\bibinfo{date}{March} \bibinfo{year}{2021}), \bibinfo{pages}{99--118}.
\newblock
\showISSN{2255-4165}
\href{https://doi.org/10.12840/ISSN.2255-4165.028}{doi:\nolinkurl{10.12840/ISSN.2255-4165.028}}


\bibitem[Turner(2022)]%
        {turner2022}
\bibfield{author}{\bibinfo{person}{Cody Turner}.} \bibinfo{year}{2022}\natexlab{}.
\newblock \showarticletitle{Augmented {Reality}, {Augmented} {Epistemology}, and the {Real}-{World} {Web}}.
\newblock \bibinfo{journal}{\emph{Philosophy \& Technology}} \bibinfo{volume}{35}, \bibinfo{number}{1} (\bibinfo{date}{March} \bibinfo{year}{2022}), \bibinfo{pages}{19}.
\newblock
\showISSN{2210-5433, 2210-5441}
\href{https://doi.org/10.1007/s13347-022-00496-5}{doi:\nolinkurl{10.1007/s13347-022-00496-5}}


\bibitem[Vesanen(2007)]%
        {vesanen2007}
\bibfield{author}{\bibinfo{person}{Jari Vesanen}.} \bibinfo{year}{2007}\natexlab{}.
\newblock \showarticletitle{What is personalization? {A} conceptual framework}.
\newblock \bibinfo{journal}{\emph{European Journal of Marketing}} \bibinfo{volume}{41}, \bibinfo{number}{5/6} (\bibinfo{date}{June} \bibinfo{year}{2007}), \bibinfo{pages}{409--418}.
\newblock
\showISSN{0309-0566}
\href{https://doi.org/10.1108/03090560710737534}{doi:\nolinkurl{10.1108/03090560710737534}}


\bibitem[Wei et~al\mbox{.}(2024)]%
        {wei2024}
\bibfield{author}{\bibinfo{person}{Qianjie Wei}, \bibinfo{person}{Jingling Zhang}, \bibinfo{person}{Pengqi Wang}, \bibinfo{person}{Xiaofu Jin}, {and} \bibinfo{person}{Mingming Fan}.} \bibinfo{year}{2024}\natexlab{}.
\newblock \showarticletitle{Augmented {Library}: {Toward} {Enriching} {Physical} {Library} {Experience} {Using} {HMD}-{Based} {Augmented} {Reality}}. In \bibinfo{booktitle}{\emph{Proceedings of the 17th {International} {Symposium} on {Visual} {Information} {Communication} and {Interaction}}} \emph{(\bibinfo{series}{{VINCI} '24})}. \bibinfo{publisher}{ACM}, \bibinfo{address}{New York, NY, USA}, \bibinfo{pages}{1--5}.
\newblock
\showISBNx{979-8-4007-0967-8}
\href{https://doi.org/10.1145/3678698.3687174}{doi:\nolinkurl{10.1145/3678698.3687174}}


\bibitem[Yang et~al\mbox{.}(2022)]%
        {yang2022a}
\bibfield{author}{\bibinfo{person}{Felix Yang}, \bibinfo{person}{Saikishore Kalloori}, \bibinfo{person}{Ribin Chalumattu}, {and} \bibinfo{person}{Markus Gross}.} \bibinfo{year}{2022}\natexlab{}.
\newblock \showarticletitle{Personalized {Information} {Retrieval} for {Touristic} {Attractions} in {Augmented} {Reality}}. In \bibinfo{booktitle}{\emph{Proceedings of the {Fifteenth} {ACM} {International} {Conference} on {Web} {Search} and {Data} {Mining}}} \emph{(\bibinfo{series}{Wsdm '22})}. \bibinfo{publisher}{Association for Computing Machinery}, \bibinfo{address}{New York, NY, USA}, \bibinfo{pages}{1613--1616}.
\newblock
\showISBNx{978-1-4503-9132-0}
\href{https://doi.org/10.1145/3488560.3502194}{doi:\nolinkurl{10.1145/3488560.3502194}}


\bibitem[Yeung(2018)]%
        {yeung2018}
\bibfield{author}{\bibinfo{person}{Karen Yeung}.} \bibinfo{year}{2018}\natexlab{}.
\newblock \showarticletitle{Five fears about mass predictive personalization in an age of surveillance capitalism}.
\newblock \bibinfo{journal}{\emph{International Data Privacy Law}} \bibinfo{volume}{8}, \bibinfo{number}{3} (\bibinfo{date}{Aug.} \bibinfo{year}{2018}), \bibinfo{pages}{258--269}.
\newblock
\showISSN{2044-3994, 2044-4001}
\href{https://doi.org/10.1093/idpl/ipy020}{doi:\nolinkurl{10.1093/idpl/ipy020}}


\bibitem[Young(2022)]%
        {young2022}
\bibfield{author}{\bibinfo{person}{Greyson~K. Young}.} \bibinfo{year}{2022}\natexlab{}.
\newblock \showarticletitle{How much is too much: the difficulties of social media content moderation}.
\newblock \bibinfo{journal}{\emph{Information \& Communications Technology Law}} \bibinfo{volume}{31}, \bibinfo{number}{1} (\bibinfo{date}{Jan.} \bibinfo{year}{2022}), \bibinfo{pages}{1--16}.
\newblock
\showISSN{1360-0834}
\href{https://doi.org/10.1080/13600834.2021.1905593}{doi:\nolinkurl{10.1080/13600834.2021.1905593}}
\newblock
\shownote{Publisher: Routledge \_eprint: https://doi.org/10.1080/13600834.2021.1905593}.


\end{thebibliography}

% \appendix

\end{document}